\documentclass[12pt]{article}
\usepackage[english]{babel}
\usepackage{amsmath}
\usepackage{amssymb}
\usepackage{bbm}
\usepackage{color}
\usepackage[titletoc,toc,title]{appendix}
\Roman{section}
\newcommand{\naw}[1]{\left(#1\right)}

\newcommand{\av}[1]{\left<#1\right>}
\newcommand{\com}[1]{\left[#1\right]}
\newcommand{\modu}[1]{\left|#1\right|}
\newcommand{\poisson}[1]{\left\{#1\right\}}

\title{Bell inequalities for $S_4$ group: classical and quantum bounds for  two orbits case}
\author{Katarzyna Bolonek-Laso\'n$^1$\footnote{katarzyna.bolonek@uni.lodz.pl}  \and Piotr Kosi\'nski$^2$\footnote{piotr.kosinski@uni.lodz.pl} }
\date{%
\small{$^1$Department of Statistical Methods, Faculty of Economics and Sociology\\ University of Lodz, 41/43 Rewolucji 1905 St., 90-214 Lodz,  Poland}\\%
$^2$Department of Computer Science, Faculty of Physics and Applied Informatics University of Lodz, 149/153 Pomorska St., 90-236 Lodz, Poland }

\begin{document}
\maketitle

\begin{abstract}
Within G\"uney-Hillery approach a number of examples of classical and quantum bounds on the sum of probabilities resulting from two orbits of $S_4$ is considered. It is shown that the violation of Bell's inequalities is rather rare and gentle.
\end{abstract}

\section{Introduction}
Since the pioneering Bell's paper \cite{Bell} the Bell inequalities became the subject of intensive study \cite{Clauser}$\div$\cite{Cabello}; for a review see \cite{Liang}, \cite{Brunner} and numerous references therein.

The basic assumption underlying the derivation of the Bell inequalities is local realism. Consequently, their violation on the level of the quantum theory provides an evidence that the latter cannot be viewed as a local realistic theory.

Recently an interesting approach to the Bell inequalities has been proposed by G\"uney and Hillery \cite{Guney}, \cite{Guney1}. It is based on the assumption that the space of states spans some representation of a finite group $G$; $G$ may be a symmetry group of the system but this assumption is not necessary.

The approach proposed in \cite{Guney}, \cite{Guney1} has been further developed in Refs. \cite{Bolonek}$\div$\cite{Yang}.
It relies on the construction of the specific set of states forming a number of orbits of $G$ in the space of states. One has to compare the classical and quantum bounds on the sum of probabilities corresponding to all points on the orbits. While computing the quantum bound is quite a straightforward exercise in representation theory, determining its classical counterpart may be a real challenge. The key problem is to find an appropriate set of orbits providing the example of Bell's inequality violation. It follows from the analysis of the examples considered so far that, typically, the violation is rather gentle.

In the present paper we continue the study of the Bell inequalities related to the standard representation of $S_4$ group. They have been already studied in Refs. \cite{Bolonek}, \cite{Bolonek1}, \cite{Bolonek2}. We restrict ourselves to the two-orbit case. In Ref. \cite{Bolonek} an example of the violation of Bell's inequalities based on $S_4$ group has been found. Our present analysis is more systematic and based on the grouptheoretical classification of states described in \cite{Bolonek2}. We find that it is quite difficult to provide further examples of the Bell inequality violation. In fact, the only case we find here is already known from Ref. \cite{Bolonek}. We didn't consider all $\binom{24}{2}=276$ nontrivial pairs of orbits. However, we analyzed a sufficient number of cases to be able to conclude that, typically, Bell's inequalities are not violated in the case under consideration, based on standard representation of $S_4$ group acting in both parties of bipartite system. 
In the last section we give some general arguments in favor of the conclusion that within the grouptheoretical approach considered here the violation of Bell's inequalities is a rather subtle effect.

The paper is organized as follows. In Sec.~II we sketch the G\"uney-Hillery formalism \cite{Guney}, \cite{Guney1} and the corresponding grouptheoretical classification of states \cite{Bolonek2}. In Sec.~III the classical and quantum bounds based on the standard representation of $S_4$ are analyzed in some detail. Sec.~IV is devoted to some conclusions. More detailed information concerning technicalities is relegated to the Appendix.

\section{Grouptheoretical framework for the Bell inequalities}
Let us start with the short recapitulation of the main ideas underlying the grouptheoretical approach to Bell inequalities proposed by G\"uney and Hillery \cite{Guney}, \cite{Guney1}. One considers a bipartite system (immortal Alice and Bob) carrying a representation of some (symmetry/dynamical)  group $G$; the representation under consideration is the tensor product of two isomorphic irreducible representations of $G$ acting in the Alice and Bob spaces, respectively. The sets of states entering Bell inequalities belong to some orbits of $G$ in the space of states. The main point is to select orbits which decompose into disjoint sets of mutually orthogonal vectors providing the orthonormal bases; the latter define spectral decompositions of some observables. In Ref. \cite{Bolonek2} the following scheme has been proposed which generalizes the examples considered in \cite{Guney1}.

Assume that $\poisson{D(g)\vert g\in G}$ is $m$ dimensional irreducible (over $\mathbb{C}$) representation of $G$ acting in the Alice (Bob) space. By the well-known theorem \cite{Simon} $m$ divides the order of $G$, $\frac{\vert G\vert}{m}=k\in \mathbb{N}$. Assume further that $G$ possesses a cyclic subgroup $H\subset G$ of order $m$, $H=\poisson{g^l\vert l=0,\ldots,m-1}$ and there exists a state vector $v$ such that: (\textit{i}) the orbit $O_v=\poisson{D(\tilde{g})v\vert \tilde{g}\in G}$ is regular (i.e. consists of exactly $\vert G\vert$ elements); (\textit{ii}) $(v, D(g^l)v)=0$ for $l=1,\ldots, m-1$. Then one easily finds that $\poisson{v, D(g)v,\ldots,D(g^{m-1})v}$ is an orthonormal basis in the Alice (or Bob) space of  states. Let $\poisson{g_1=e,g_2,\ldots,g_k}$ be a set of representatives of left cosets from $G/H$; any $\tilde{g}\in G$ can be written as
\begin{equation}
\tilde{g}=g_\alpha g^l,\quad \alpha=1,\ldots,k,\quad l=0,\ldots,m-1.
\end{equation}
Define
\begin{equation}
v_{\alpha l}\equiv D(\tilde{g})v=D(g_\alpha)D(g^l)v.
\end{equation}
Then the vectors $v_{\alpha l}$, $\alpha$-fixed, $l=0,\ldots,m-1$, form an orthonormal basis. By definition they provide the spectral decomposition of some Alice (Bob) observable $A_\alpha$ ($B_\alpha$). Therefore, the orbit $O_v$ decomposes into disjoint orthonormal bases defining the observables $A_1,\,A_2,\ldots,\,A_k$ ($B_1,\,B_2,\ldots,\,B_k$). For a complete definition of $A_\alpha$'s one should specify the corresponding eigenvalues; this is, however, irrelevant for what follows because the Bell inequalities can be formulated in terms of probabilities alone. \\
Similar construction is performed in the Bob space of states. The next step is to define an orbit in the total space of the bipartite system. We shall consider the following special class of orbits:
\begin{equation}
O(\tilde{g},v)=\poisson{D(g')v\otimes D(g'\tilde{g})v\vert g'\in G}
\end{equation}
where $\tilde{g}\in G$ is some fixed element while $v$ has been specified above. Note that writting
\begin{equation}
g'=g_\alpha g^l
\end{equation}
one obtains 
\begin{equation}
g'\tilde{g}=g_{\alpha_{\tilde{g}}}g^{l_{\tilde{g}}}
\end{equation}
where
\begin{equation}
(\alpha,l)\rightarrow(\alpha,l)_{\tilde{g}}\equiv (\alpha_{\tilde{g}},l_{\tilde{g}})
\end{equation}
is some permutation of $\vert G\vert$ pairs $(\alpha,l)$. Accordingly, $O(\tilde{g},v)$ takes the following form
\begin{equation}
O(\tilde{g},v)=\poisson{v_{\alpha l}\otimes v_{\alpha_{\tilde{g}}l_{\tilde{g}}}\vert \,\alpha=1,\ldots,k,\, l=0,\ldots,m-1}.
\end{equation}
Let now $w$ be some normalized state of the total system; then 
\begin{equation}
p(\alpha,l;\alpha_{\tilde{g}},l_{\tilde{g}})=\modu{\naw{v_{\alpha l}\otimes v_{\alpha_{\tilde{g}}l_{\tilde{g}}},w}}^2
\end{equation}
is the probability that, for the system in the state described by $w$, the simultaneous mesurement of the observables $A_{\alpha}$ (Alice) and $B_{\alpha_{\tilde{g}}}$ (Bob) yields the values corresponding to $l$-th and $l_{\tilde{g}}$-th eigenvectors, respectively. In order to estimate the sum 
\begin{equation}
S=\sum\limits_{(\alpha,l)}p(\alpha,l;\alpha_{\tilde{g}},l_{\tilde{g}})\label{a9}
\end{equation}
one notices that $S$ can be written as
\begin{equation}
S=(w,Xw)
\end{equation}
where 
\begin{equation}
X\equiv X(\tilde{g},v)=\sum\limits_{g'\in G}\naw{D(g')v\otimes D(g')D(\tilde{g})v}\naw{D(g')v\otimes D(g')D(\tilde{g})v}^+\label{a11}
\end{equation}
$S$ is bounded from above by the maximal eigenvalue of $X(\tilde{g},v)$. It is easy to check that $X(\tilde{g},v)$ commutes with all $D(g')\otimes D(g')$, $g'\in G$. Now, $D\otimes D$ is, in general, reducible and decomposes into the direct sum of irreducible components,
\begin{equation}
D\otimes D=\bigoplus_s D_s.\label{a12}
\end{equation}
Assuming that each $D_s$ appears on the right hand side of (\ref{a12}) at most once, one concludes that $X(\tilde{g},v)$ is diagonal in the basis in which the decomposition (\ref{a12}) is explicit; moreover, $X(\tilde{g},v)$ reduces to a multiplicity of unity on any irreducible component. The eigenvalues of $X(\tilde{g},v)$ are obtained from orthogonality relations and read \cite{Guney1}
\begin{equation}
x_s=\frac{\vert G\vert}{d_s}\Vert\naw{v\otimes D(\tilde{g})v}_s\Vert^2\label{a13}
\end{equation}
where $d_s$ is the dimension of $D_s$ while $\naw{v\otimes D(\tilde{g})v}_s$ - the projection of $v\otimes D(\tilde{g})v$ onto the carrier space of $D_s$. Therefore, one finds the estimate 
\begin{equation}
S\leq \max\limits_{s}\frac{\vert G\vert}{d_s}\Vert\naw{v\otimes D(\tilde{g})v}_s\Vert^2\equiv \frac{\vert G\vert}{d_{s_0}}\Vert\naw{v\otimes D(\tilde{g})v}_{s_0}\Vert^2.\label{a14}
\end{equation}
The upper bound is attained for any vector $w$ belonging to the subspace carrying the representation $D_{s_0}$.

Eq. (\ref{a14}) provides the quantum upper bound on the sum of probabilities (\ref{a9}); it could be vaguely called Tsirelson-like bound \cite{Cirelson}. Now, we would like to study the classical bound on $S$;
the latter constitutes the Bell inequality. All $A_{\alpha}$'s commute with all $B_{\alpha}$'s so the sum (\ref{a9}) makes sense both on the classical and quantum levels. However, due to the fact that, in general, $\com{A_\alpha,A_{\alpha'}}\neq 0$, $\com{B_\alpha,B_{\alpha'}}\neq 0$, the existence of joint probability distribution
\begin{equation}
p(\underline{a};\underline{b})=p(A_1=a_1,\ldots,A_k=a_k;\, B_1=b_1,\ldots, B_k=b_k)\label{a15}
\end{equation}
is allowed only on the classical level. Then the probabilities entering (\ref{a9}) are returned as appropriate marginals. Inserting the relevant expression into the right hand side of (\ref{a9}) one finds \cite{Guney}$\div$\cite{Bolonek2}
\begin{equation}
S=\sum_{(\underline{a},\underline{b})}c(\underline{a},\underline{b})p(\underline{a},\underline{b})\label{a16}
\end{equation}
where the sum runs over all configurations $(\underline{a},\underline{b})\equiv(a_1,\ldots,\,a_k;\, b_1,\ldots,b_k)$ and nonnegative integers $c(\underline{a},\underline{b})$ indicate the number of times a particular configuration $(\underline{a},\underline{b})$ enters the sum defining $S$. Due to $0\leq p(\underline{a},\underline{b})\leq 1$, $\sum\limits_{(\underline{a},\underline{b})}p(\underline{a},\underline{b})=1$, eq. (\ref{a16}) implies 
\begin{equation}
S\leq\max_{(\underline{a},\underline{b})}c(\underline{a},\underline{b})\label{a17}
\end{equation}
Eq. (\ref{a17}) is our Bell inequality. Note that one can always select the  joint probability maximizing $S$ in the form 
\begin{equation}
p(\underline{a},\underline{b})=\left\{\begin{array}{ll}
1 & \text{if}\; (\underline{a},\underline{b})=(\underline{a}^{(0)},\underline{b}^{(0)})\\
0 & \text{otherwise}
\end{array}\right..\label{a18}
\end{equation}
Once an appropriate orbit is selected one computes from (\ref{a14}) and (\ref{a17}) the quantum and classical bounds, respectively. If the former exceeds the latter, the Bell inequality is violated.

The above reasoning can be carried out for few orbits $O(\tilde{g}_1,v),\ldots,O(\tilde{g}_r,v)$. On the quantum level the corresponding operators $X(\tilde{g}_a,v)$, $a=1,\ldots,r$, commute so they are simultaneously diagonalizable and one has only to select the maximal sum of relevant  eigenvalues. 
Thus the grouptheoretical approach provides an algorithm allowing to determine the quantum upper bound on the sum $S$. On the other hand, eqs. (\ref{a16})$\div$(\ref{a18}) suggest that the task of finding the classical bound on $S$ is a combinatorial problem: given a number $r$ of orbits $O(\tilde{g}_a,v)$ characterized by the elements $\tilde{g}_1,\,\tilde{g}_2,\dots,\,\tilde{g}_r$ what is the upper bound on the sum 
\begin{equation}
S(\tilde{g}_1,\ldots,\tilde{g}_r)\equiv\sum_{a=1}^r\sum_{(\alpha,l)}p(\alpha,l;\alpha_{\tilde{g}_a},l_{\tilde{g}_a})\label{ab19}
\end{equation}
of classical (i.e. obtained as marginals from joint distribution) probabilities; it equals the maximal number of times some joint configuration enters the sum (\ref{ab19}).

The problem can be easily solved in the case of one orbit \cite{Bolonek2}. First, let us note that, due to the fact that the bound on $S$ is saturated by the joint probability (\ref{a18}) concentrated on a single configuration, $S$ is bounded by $k$,  the number of elements in $G/H$. In order to show that the bound is actually attained one can apply  Hall marriage theorem \cite{Cameron}. This result suggests that it is unlikely to find the violation of the  Bell inequality on quantum level if only one orbit is taken into account \cite{Bolonek2}.

On the other hand the known examples \cite{Guney}$\div$\cite{Bolonek2} show that for two and three orbits the Bell inequalities can be violated; however, the relevant orbits must be carefully chosen.

Consider the case of two orbits, $r=2$. An a priori classical upper bound on the sum (\ref{a9}) varies between $k$ and $2k$. In some cases it is quite easy to identify the pairs of orbits with a given classical upper bound. The simplest example is provided by the pairs corresponding to the lowest value $k$ of the upper bound. Again, we use the property that the upper bound is attained for a joint probability distribution concentrated on one configuration. Let $(\alpha,l(\alpha))$, $\alpha=1,2,\ldots,k$, be the elements of Alice orbit belonging to this configuration. The corresponding Bob's elements are $(\alpha_{\tilde{g}_1},l(\alpha)_{\tilde{g}_1})$ and $(\alpha_{\tilde{g}_2},l(\alpha)_{\tilde{g}_2})$ for the first and second orbits, respectively. If the following condition holds:  $l(\alpha)_{\tilde{g}_1}\neq l(\alpha')_{\tilde{g}_2}$ provided $\alpha_{\tilde{g}_1}=\alpha'_{\tilde{g}_2}$, $\alpha=1,\ldots,k$ then only $k$ elements from both orbits can belong to a given configuration; in other words, $c(\underline{a},\underline{b})\leq k$. It is not difficult to arrange such a situation. Namely, let $\tilde{g}_1^{-1}\tilde{g}_2\in H$, $\tilde{g}_1\neq\tilde{g}_2$, i.e.
\begin{equation}
\tilde{g}_2=\tilde{g}_1g^r,\quad r=1,\ldots,m-1;
\end{equation}  
then 
\begin{equation}
\alpha_{\tilde{g}_1}=\alpha_{\tilde{g}_2},\quad l_{\tilde{g}_2}=l_{\tilde{g}_1}+r(\text{mod}\, m).
\end{equation}
However, the lowest possible value $k$ of the classical bound can be attained even if $\tilde{g}_1^{-1}\tilde{g}_2\notin H$ as shown by the examples considered in the next section. The situation becomes even more involved if we are interested in the bounds
\begin{equation}
S(\tilde{g}_1,\tilde{g}_2)\leq B
\end{equation}
obeying $k<B\leq 2k$. The method of analyzing such cases has been developed in Ref. \cite{Bolonek}. Strictly speaking, it allows us to check that $B$ is not smaller than some number $B'$ which can be quite easily determined; showing that actually $B$ equals $B'$ demands separate proof. However, if we already know that the quantum bound is not larger than $B'$, the Bell inequality cannot be violated. This is the case for most examples considered in the next section; in the single case of the Bell inequality violation the proof that $B=B'$ has been already given in \cite{Bolonek}. The method proposed in \cite{Bolonek} is briefly sketched in Appendix using grouptheoretical notation. 

\section{Bounds for symmetric group $S_4$}
We shall consider now the case of the standard representation of symmetric group $S_4$. Some particular examples of Bell's inequalities and their violation has been already discussed in \cite{Bolonek} and \cite{Bolonek1}. Here we present a more detailed discussion based on the general construction of states described in \cite{Bolonek} and sketched in previous section. 

$S_4$ has five inequivalent irreducible representations: trivial ($D_0$), alternating ($\tilde{D}_0$), twodimensional ($D_2$), threedimensional standard representation ($D$) and another threedimensional one ($\tilde{D}$). The latter is actually the product of $D$ and $\tilde{D}_0$. All representations are real and only threedimensional ones are faithful. The standard representation is obtained by considering the natural action of $S_4$ in fourdimensional space and deleting the invariant subspace $x_1+x_2+x_3+x_4=0$. Its explicit form in unitary (orthogonal) basis is described in Appendix. Some appropriate orbits have been constructed in \cite{Bolonek}, \cite{Bolonek1} by the purely geometric means; essentially, using the fact that $S_4$ is the symmetry group of regular tetrahedron. However, in order to keep the discussion more general, we follow the algorithm outlined above. We take $H$ to be the cyclic subgroup of order $m=3$, $H=\{e,g,g^2\}$, generated by $g=(2314)$. One easily finds that this choice, together with 
\begin{equation}
v=\frac{1}{\sqrt{3}}\naw{\begin{array}{c}
1\\1\\1
\end{array}},\label{a22}
\end{equation} 
is consistent with the assumptions (\textit{i}) and (\textit{ii}) of the previous section. There are $k=8$ Alice (Bob) observables, in one-to-one correspondence with the elements of $G/H$. The representatives of the corresponding cosets, $g_\alpha$, $\alpha=1,\ldots,8$, are given in Table \ref{t1}  in the Appendix. The eigenvalues of the operators $X(\tilde{g},v)$ for all 24 orbits in the total space of bipartite system are computed according to the prescription described in Sec. II. They are collected in the Table \ref{t2} from the Appendix. We see that the maximal eigenvalue equals 8 and is attained for singlet state; this is to be expected because this state is maximally entangled. However, for one orbit the classical bound is also equal to 8 as follows from Hall matching theorem \cite{Cameron}. Therefore, as expected \cite{Guney}$\div$\cite{Bolonek2}, Bell's inequality is not violated on the one orbit level.

In what follows we will be interested in the two orbit case. The corresponding operator $X$ reads
\begin{equation}
X=X(\tilde{g}_1,v)+X(\tilde{g}_2,v)
\end{equation}
where $\tilde{g}_1,\tilde{g}_2\in S_4$ are arbitrary group elements. Using Table \ref{t2} we can find the eigenvalues of $X$ for all pairs $\tilde{g}_1,\,\tilde{g}_2$; this provides the quantum bound which equals to the largest eigenvalue.

In order to find the classical bounds we use the method described in the Appendix. As it has been discussed in the previous section the classical bound (Bell's inequality) is given by an integer $B$ obeying $k\leq B\leq 2k$; in the case under consideration $k=8$. We found that $B=8,\,12,\,14,\,16$; we were unable to construct an example corresponding to $B=10$. Below we discuss the particular cases providing the examples of Bell's inequalities including the case the relevant inequality is violated. 

\underline{$B=8$}\\
In Sec. II it has been shown that the inequality
\begin{equation}
S(\tilde{g}_1,\tilde{g}_2)\leq k \quad (=8)\label{ab1}
\end{equation}
holds provided $\tilde{g}_1^{-1}\tilde{g}_2\in H$, $\tilde{g}_1\neq\tilde{g}_2$. This implies that the orbits under consideration, $O(\tilde{g}_1,v)$, $O(\tilde{g}_2,v)$, consist of the vectors $v\otimes v_{\alpha l}$  and $v\otimes v_{\alpha l'}$, $l\neq l'$, respectively. By inspecting Table \ref{t2} we see that Bell's inequality is not violated for all $\alpha=1,\ldots,8$, $l\neq l'=0,1,2$.

 The condition $\tilde{g}_1^{-1}\tilde{g}_2\in H$ is sufficient but not necessary for (\ref{ab1}) to hold. One can find another examples of pairs of orbits with $S(\tilde{g}_1,\tilde{g}_2)$ obeying (\ref{ab1}). In particular, we considered the pairs of orbits containing the vectors: $(v\otimes v_{42}, v\otimes v_{70})$, $(v\otimes v_{32},v\otimes v_{41})$, $(v\otimes v_{31},v\otimes v_{71})$, $(v\otimes v_{50},v\otimes v_{80})$, $(v\otimes v_{60},v\otimes v_{81})$, $(v\otimes v_{41},v\otimes v_{72})$, $(v\otimes v_{32},v\otimes v_{72})$, $(v\otimes v_{40},v\otimes v_{71})$ and $(v\otimes v_{30},v\otimes v_{42})$. In no case we found that the Bell inequality is violated. 
 
 In what follows we shall identify, as above, the pairs of orbits with the pairs of vectors $(v\otimes \tilde{g}_1v,v\otimes\tilde{g}_2v)$.

\underline{$B=10$}\\
As it has been mentioned above we were not able to find examples of pairs of orbits corresponding to $B=10$. We didn't check all pairs of orbits but it seems plausible that no such pairs exist. However, our final aim is to find the examples of the violation of the Bell inequalities. Therefore, we have selected, using the Table 2 from Appendix, all pairs of orbits for which the violation is potentially possible. Then we have checked no such pair yields the classical bound $B=10$.  

\underline{$B=12$}\\
We considered the following examples of pairs of orbits corresponding to the classical bound $B=12$: $(v\otimes v_{40},v\otimes v_{72})$, $(v\otimes v_{30},v\otimes v_{72})$, $(v\otimes v_{30},v\otimes v_{40})$, $(v\otimes v_{41},v\otimes v_{70})$, $(v\otimes v_{31},v\otimes v_{70})$, $(v\otimes v_{31},v\otimes v_{41})$, $(v\otimes v_{42},v\otimes v_{71})$, $(v\otimes v_{32},v\otimes v_{71})$, $(v\otimes v_{32},v\otimes v_{42})$ and $(v\otimes v_{52},v\otimes v_{81})$. Again no example of the Bell inequality violation has been found. 

\underline{$B=14$}\\
The following pairs of orbits have been analyzed: $(v\otimes v_{22},v\otimes v_{72})$, $(v\otimes v_{21},v\otimes v_{72})$, $(v\otimes v_{22},v\otimes v_{40})$, $(v\otimes v_{20},v\otimes v_{40})$, $(v\otimes v_{20},v\otimes v_{30})$, $(v\otimes v_{21},v\otimes v_{30})$, $(v\otimes v_{21},v\otimes v_{42})$, $(v\otimes v_{22},v\otimes v_{32})$, $(v\otimes v_{20},v\otimes v_{41})$, $(v\otimes v_{21},v\otimes v_{31})$, $(v\otimes v_{20},v\otimes v_{32})$, $(v\otimes v_{31},v\otimes v_{50})$. We found only one example of violation of the Bell inequality. This is the pair of orbits  corresponding to the vectors $(v\otimes v_{22}, v\otimes v_{72})$; the quantum bound equals 14.036 and corresponds, as expected, to maximally entangled vector transforming according to the trivial representation $D_0$. This example has been already considered in \cite{Bolonek}.

\underline{$B=16$}\\
We found several examples of pairs of orbits corresponding to this value of classical bound. However, all eigenvalues of $X(\tilde{g},v)$, corresponding to single orbits, are bounded by 8. Therefore, no violation of the Bell inequality is possible. 

\section{Conclusions}

We have studied in some detail the G\"uney-Hillery approach to the Bell inequalities for the standard representation of the symmetry group $S_4$. $S_4$ possesses five irreducible representations.  The onedimensional representations (trivial and sign ones) are not interesting because then all observables commute. The twodimensional representation is homomorphic; the quotient group of $S_4$ by its kernel is isomorphic to $S_3$. Therefore, twodimensional representation is the faithful representations of $S_3$. In the context of G\"uney-Hillery scheme this case has been already considered in Ref. \cite{Guney1}. Both threedimensional representations are faithful so they fully reflect the structure of $S_4$. However, the second representation is obtained from the standard one by multiplying the matrices of the latter by the signs of the corresponding permutations; consequently, our conclusions can be easily extended to this case. Therefore, it is justified to consider our example as reflecting the properties of $S_4$ which are relevant in G\"uney-Hillery framework.

The main conclusions of our study are:
\begin{itemize}
\item[(\textit{i})] in the present context the violation of Bell inequalities is a rather rare phenomenon; we haven't analyzed all $\binom{24}{2}=276$ possibilities but the number of cases we considered is significant enough to draw such conclusion;
\item [(\textit{ii})] the violation, if it appears, is rather gentle; in the single example we found it is of order $0.25\%$.
\end{itemize}
As it has been already explained the two-orbit case is the simplest case where the violation of Bell's inequalities can occur. Let us compare our findings with the results obtained in the $S_3$ case. It has been found in Ref. \cite{Guney1} that then a particular choice of two orbits leads to the violation of the Bell inequality by $5\%$ which is significantly stronger than in the $S_4$ case. This result concerns again the standard representation of $S_3$ which is the only irreducible representation apart from trivial and sign ones. Let us also note that the example found in \cite{Guney1} is essentially unique as can be shown by repeating the reasoning outlined in the present paper.

These results should be compared with those concerning basic CHSH inequality for four dichotomic variables. The classical bound equals here 2 \cite{Clauser} while the quantum reads $2\sqrt{2}$ \cite{Cirelson}; both bounds are saturated. 
Therefore, the Bell inequality is violated in about $41\%$. This is strong violation as compared with the one considered here. One can argue that the former concerns the expectation values and not the probabilities themseleves; however, the eigenvalues of the observables under consideration equal $\pm 1$ so the order of magnitude is correct.

In order to see the origin of the difference between the CHSH-type inequalities and those considered here let us note that the probability sums we are considering here (cf. eq.~(\ref{a9})) are quite special. First, all probabilities enter with positive (actually, $+1$) coefficients; moreover, any configuration of one party is paired with exactly one configuration of the second party. These properties are not shared by the expressions entering general Bell's inequalities. As an example let us consider simple CHSH inequality mentioned above
\begin{equation}
\modu{\av{A_0B_0}+\av{A_0B_1}+\av{A_1B_0}-\av{A_1B_1}}\leq 2\label{25a}
\end{equation}    
with $A_{\alpha}=\pm 1$, $B_\beta=\pm 1$, $\alpha,\beta=0,1$. Using 
\begin{equation}
\begin{split}
& \av{A_\alpha B_\beta}=p\naw{A_\alpha=1;B_\beta=1}+p(A_\alpha=-1;B_\beta=-1)\\
& \hspace{1.5 cm} -p(A_\alpha=1;B_\beta=-1)-p(A_\alpha=-1;B_\beta=1)
\end{split}\label{25b}
\end{equation}
one represents the left hand side of (\ref{25b}) as a linear combination of probabilities with both positive and negative coefficients. In terms of joint probability distribution $p(A_0=a_0, A_1=a_1;B_0=b_0,B_1=b_1)$ it reads
\begin{equation}
\begin{split}
& 2\left({\sum_{\substack{a_\alpha=\pm 1\\ b_\beta=\pm 1}}}^{'}p(A_0=a_0,A_1=a_1;B_0=b_0,B_1=b_1)\right.\\
&\left.-{\sum_{\substack{a_\alpha=\pm 1\\ b_\beta=\pm 1}}}^{''}p(A_0=a_0,A_1=a_1;B_0=b_0,B_1=b_1)\right)
\end{split}\label{25c}
\end{equation}
where both sums run over disjoint subsets of configurations. (\ref{25c}) yields an immediate proof of the CHSH inequality. Let us note that one can redefine observables in such a way that all probabilities enter with positive coefficients. However, then the counterpart of eq.~(\ref{a9}) does not obey the condition that the configurations of both parties are uniquely paired.

We see that the grouptheoretical approach refers to some particular combinations of probabilities. Therefore, the violation of the Bell inequalities, obtained within grouptheoretical framework considered here, are expected to be weaker than in the general case. Let us note that the specific combination of the correlation functions entering the left hand side of eq.~(\ref{25a}) is chosen in order to provide a number of cancellations resulting in the nice final expression (\ref{25c}). The grouptheoretical analysis shows that the violation of the Bell inequalities is also possible if such a cancellation does not take place. Moreover, we are convinced that the G\"uney-Hillery approach can be extended to more general expressions as those entering CHSH-like inequalities. We hope to return to this problem in future publication.

Another interesting problem is the efficiency  of G\"uney-Hillery approach for other groups and representations. Obviously, one should consider mainly faithful representations; the nonfaithful ones reflect the properties  of quotient groups. Moreover, roughly speaking one expects the stronger violation the more entangled the quantum state is. If both parties transform according to the same (pseudo) real representation $D$ then the singlet state contained in $D\otimes D$ is maximally entangled; this indicates that then one should look for the examples of violation among singlet states. The natural generalization of the analysis presented here is to consider the standard representation of $S_n$ group. Our conjecture is that with growing $n$ the degree of violation of the Bell inequalities decreases. 

\appendix
\section{Appendix}
\underline{The explicit form of standard representation of $S_4$}\\

Below we present the explicit form of standard representation of $S_4$ in an unitary basis. To this end it is sufficient to know the matrices representing transpositions. They read \cite{Bolonek}:
\begin{equation}
\begin{split}
& D\naw{12}=\left[\begin{array}{ccc}
1 & 0 & 0\\
0 & 1 & 0\\
0 & 0 & -1
\end{array}\right],\qquad\qquad\quad
 D\naw{13}=\left[\begin{array}{ccc}
1 & 0 & 0\\
0 & -\frac{1}{2} & -\frac{\sqrt{3}}{2}\\
0 & -\frac{\sqrt{3}}{2} & \frac{1}{2}
\end{array}\right]\\
& D\naw{14}=\left[\begin{array}{ccc}
-\frac{1}{3} & -\frac{\sqrt{2}}{3} & -\frac{\sqrt{6}}{3}\\
-\frac{\sqrt{2}}{3} & \frac{5}{6} & -\frac{\sqrt{3}}{6}\\
-\frac{\sqrt{6}}{3} & -\frac{\sqrt{3}}{6}& \frac{1}{2}
\end{array}\right],\qquad
 D\naw{23}=\left[\begin{array}{ccc}
1 & 0 & 0\\
0 & -\frac{1}{2} & \frac{\sqrt{3}}{2}\\
0 & \frac{\sqrt{3}}{2} & \frac{1}{2}
\end{array}\right]\\
& D\naw{24}=\left[\begin{array}{ccc}
-\frac{1}{3} & -\frac{\sqrt{2}}{3} & \frac{\sqrt{6}}{3}\\
-\frac{\sqrt{2}}{3} & \frac{5}{6} & \frac{\sqrt{3}}{6}\\
\frac{\sqrt{6}}{3} & \frac{\sqrt{3}}{6}& \frac{1}{2}
\end{array}\right],\qquad\quad
 D\naw{34}=\left[\begin{array}{ccc}
-\frac{1}{3} & \frac{\sqrt{8}}{3} & 0\\
\frac{\sqrt{8}}{3} & \frac{1}{3} & 0\\
0 & 0 & 1
\end{array}\right].\end{split}
\end{equation}
\vspace{0.2cm}

\underline{The specification of orbits}\\

We consider the orbits of $S_4$ in the threedimensional space carrying the standard representation. To this end we use the general scheme outlined in Sec. II. The cyclic subgroup $H=\poisson{e,g,g^2}$ is generated by $g=(2314)$; therefore, $m=3$ and $k=8$. In order to simplify notation we put below $D(\tilde{g})\equiv \tilde{g}$ for any $\tilde{g}\in S_4$. The initial vector $v$ is given by eq.~(\ref{a22}); using the explicit form of the representation one finds that $g^lv$, $l=0,1,2$, are mutually orthogonal. The elements of the orbit under consideration are of the form
\begin{equation}
v_{\alpha l}\equiv g_\alpha g^l v, \quad \alpha=1,\ldots, 8,\quad l=0,1,2;
\end{equation}
It remains to select the elements $g_\alpha$ representing the left cosets from $S_4/H$. The particular choice adopted here is described in Table \ref{t1}.

\begin{table}
\caption{The coset representatives}
\begin{tabular}{|c|c|c|c|c|c|c|c|c|}
\hline
$\alpha$ &1&2&3&4&5&6&7&8\\
\hline
$g_\alpha$ &e& (2134) & (4231) & (1432) & (1342) & (1423) & (2341) & (4132)\\
\hline
\end{tabular}\label{t1}
\end{table}

\underline{The eigenvalues of $X(\tilde{g},v)$}\\

The eigenvalues of the operators $X(\tilde{g},v)$ (cf. eq.~(\ref{a11})) are given  by eq.~(\ref{a13}). To find their actual values one has only to know the projections $(v_A\otimes v_B)_s$ of the product vectors on the subspaces carrying irreducible representations entering the decomposition (\ref{a12}). To this end one should compute the matrix of the relevant Clebsh-Gordan coefficients; this is quite straightforward and the final result reads \cite{Bolonek}
\begin{equation}
C=\left [\begin{array}{ccccccccc}
\sqrt{\frac{2}{3}} & 0 & 0 & 0 & -\frac{1}{\sqrt{6}} & 0 & 0 & 0 & -\frac{1}{\sqrt{6}}\\
0 & -\frac{1}{\sqrt{6}} & 0 & -\frac{1}{\sqrt{6}} & \frac{1}{\sqrt{3}} & 0 & 0 & 0 & -\frac{1}{\sqrt{3}}\\
0 & 0 & -\frac{1}{\sqrt{6}} & 0 & 0 & -\frac{1}{\sqrt{3}} & -\frac{1}{\sqrt{6}} & -\frac{1}{\sqrt{3}} & 0\\
0 & \frac{1}{\sqrt{2}} & 0 & -\frac{1}{\sqrt{2}} & 0 & 0 & 0 & 0 & 0\\
0 & 0 & \frac{1}{\sqrt{2}} & 0 & 0 & 0 & -\frac{1}{\sqrt{2}} & 0 & 0\\
 0 & 0 & 0 & 0 & 0 & \frac{1}{\sqrt{2}} & 0 & -\frac{1}{\sqrt{2}} &  0\\
 0 & \frac{1}{\sqrt{3}} & 0 & \frac{1}{\sqrt{3}} & \frac{1}{\sqrt{6}} & 0 & 0 & 0 & -\frac{1}{\sqrt{6}}\\
 0 & 0 & \frac{1}{\sqrt{3}} & 0 & 0 & -\frac{1}{\sqrt{6}} & \frac{1}{\sqrt{3}} & -\frac{1}{\sqrt{6}} & 0\\
 \frac{1}{\sqrt{3}} & 0 & 0 & 0 & \frac{1}{\sqrt{3}} & 0 & 0 & 0 & \frac{1}{\sqrt{3}}
\end{array}\right]\begin{array}{c}\left.\begin{array}{c}
\\
\\
\\
\end{array}\right\}\\
\left.\begin{array}{c}
\\
\\
\\
\end{array}\right\}\\
\left.\begin{array}{c}
\\
\\
\end{array}\right\}\\
\left.\begin{array}{c}
\\
\end{array}\right\}
\end{array}\begin{array}{c}\begin{array}{c}
\\
D \\
\\
\end{array}\\
\begin{array}{c}
\\
\tilde{D}\\
\\
\end{array}\\
\begin{array}{c}
D_2\\
\\
\end{array}\\
\begin{array}{c}
D_0
\end{array}
\end{array}
\end{equation}
The rows of $C$ correspond to the consecutive basic vectors of block-diagonal basis while the columns - to the product vectors. More explicitly, the relevant projections of $\vec{v}\otimes\vec{v}'$ onto the irreducible subspaces read \cite{Bolonek}:
\begin{equation}
\begin{split}
&D_0:\quad \frac{1}{\sqrt{3}}\naw{\vec{v}\cdot\vec{v}'}\\
& D_2:\quad \frac{1}{\sqrt{3}}\naw{v_1v_3'+v_3v_1'}-\frac{1}{\sqrt{6}}\naw{v_2v_3'+v_3v_2'}\\
& \qquad \quad \frac{1}{\sqrt{3}}\naw{v_1v_2'+v_2v_1'}+\frac{1}{\sqrt{6}}\naw{v_2v_2'-v_3v_3'}\\
&\widetilde{D}:\quad \frac{1}{\sqrt{2}}\naw{\vec{v}\times\vec{v}'}\\
& D: \quad \sqrt{\frac{2}{3}}v_1v_1'-\frac{1}{\sqrt{6}}\naw{v_2v_2'+v_3v_3'}\\
& \qquad\quad \frac{1}{\sqrt{3}}\naw{v_2v_2'-v_3v_3'}-\frac{1}{\sqrt{6}}\naw{v_1v_2'+v_2v_1'}\\
& \qquad\quad -\frac{1}{\sqrt{3}}\naw{v_2v_3'+v_3v_2'}-\frac{1}{\sqrt{6}}\naw{v_1v_3'+v_3v_1'}.
\end{split}
\end{equation}
 The eigenvalues of $X(\tilde{g},v)$, for $v$ given by eq. (\ref{a22}), are presented in Table \ref{t2}.  For convenience all elements $\tilde{g}\in S_4$ are grouped into equivalence classes. In the following two columns the elements of $S_4$ are written out together with their $(\alpha,l)$ indices; then the eigenvalues are presented which correspond to the successive irreducible components entering the Clebsch-Gordan decomposition of $D\otimes D$; finally, in the last column the maximal eigenvalues are given.

 In order to obtain the classical estimate on the sum (\ref{a16}) we may use the algorithm described in Ref. \cite{Bolonek}. We use the property that the joint probability maximizing $S$ can be chosen in the form described by eq. (\ref{a17}). Therefore, for any joint configuration $(\underline{a},\underline{b})=(a_1, \ldots,a_8;b_1,\ldots,b_8)$ (cf. eq. (\ref{a15})) we have to determine the number of times it appears in the configurations by vectors entering both orbits. To this end we start with some element $(\alpha,l)$ viewed as an Alice state. We select from the first orbit the corresponding Bob's element $(\alpha_{\tilde{g}_1},l_{\tilde{g}_1})$; then we look for the element of the second orbit containing $(\alpha_{\tilde{g}_1},l_{\tilde{g}_1})$ as a second factor.
Its first factor serves then for the search of an appropriate Alice's element of the first orbit and the procedure is repeated. As a result we obtain a closed cycle (see below). Then we select any element which does not belong to the cycle and repeat all steps. We arrive at the disjoint set of cycles of the same length and all 48 elements of both orbits belong to some cycle. Having this decomposition at hand we have to select the maximal set of cycles such that the corresponding set of vertices has the following property: for any $\alpha$, $\alpha'$ it contains at most one vertex $v_{\alpha l}\otimes v_{\alpha' l'}$. The total number of edges gives the classical bound on the sum $S(\tilde{g}_1,\tilde{g}_2)$. 
 
 In grouptheoretical language we start with the first orbit and select some element $g_0v\otimes g_0\tilde{g}_1v$; then we look for the element of the second orbit containing $g_0\tilde{g}_1v$ as its second factor. It reads  $g_0\tilde{g}_1\tilde{g}_2^{-1}v\otimes g_0\tilde{g}_1\tilde{g}_2^{-1}\tilde{g}_2v$. Coming back to the first orbit we find the vector with the same first factor of tensor product,  $g_0\tilde{g}_1\tilde{g}_2^{-1}v\otimes g_0\tilde{g}_1\tilde{g}_2^{-1}\tilde{g}_1v$ and repeat the  procedure. In this way we arrive at the sequence of vectors of the form
 \begin{equation}
 g_0(\tilde{g}_1\tilde{g}_2^{-1})^{k}v\otimes g_0(\tilde{g}_1\tilde{g}_2^{-1})^{k}\tilde{g}_1v, \quad k=0,1,\ldots\label{a28}
 \end{equation}
 from the first orbit and
  \begin{equation}
 g_0(\tilde{g}_1\tilde{g}_2^{-1})^{k+1}v\otimes g_0(\tilde{g}_1\tilde{g}_2^{-1})^{k+1}\tilde{g}_2v, \quad k=0,1,\ldots\label{a29}
 \end{equation}
 from the second one. These cycle doses if $(\tilde{g}_1\tilde{g}_2^{-1})^k = e$; therefore, for the length of the cycle (the number of vertices or edges) equals twice the order of $\tilde{g}_1\tilde{g}_2^{-1}$.
 
 Below we give the examples of this algorithm for $B=8,\,12,\,14$ and 16.\\
 
 \underline{$B=8$}: the pair of orbits containing the vectors $(v\otimes v_{42},v\otimes v_{70})$
 \setlength{\unitlength}{0.8cm}
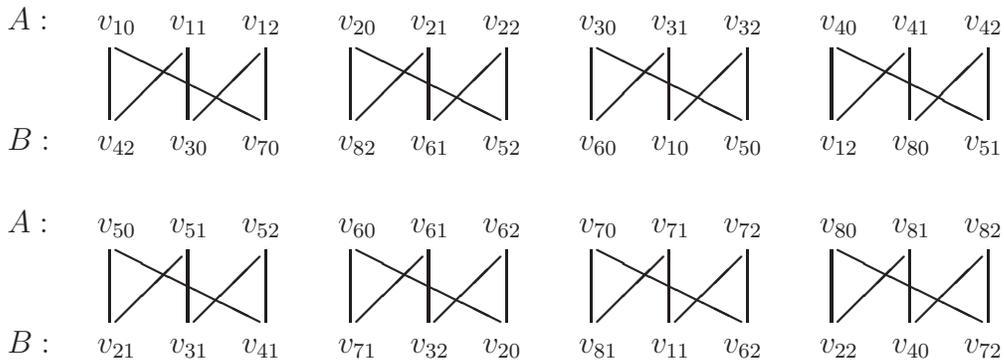
$$
\begin{picture}(4,3)
\thicklines
\put(0.2,0.5){\line(0,1){1.2}}
\put(1.5,0.5){\line(0,1){1.2}}
\put(2.8,0.5){\line(0,1){1.2}}
\put(0.3,0.5){\line(1,1){1.1}}
\put(1.6,0.5){\line(1,1){1.1}}
\put(0.3,1.7){\line(2,-1){2.4}}
\put(-1.5,2){$A:$}
\put(-1.5,0){$B:$}
\put(0,0){$v_{42}$}
\put(1.2,0){$v_{30}$}
\put(2.4,0){$v_{70}$}
\put(0,2){$v_{10}$}
\put(1.2,2){$v_{11}$}
\put(2.4,2){$v_{12}$}
\end{picture}  
\begin{picture}(4,3)
\thicklines
\put(0.2,0.5){\line(0,1){1.2}}
\put(1.5,0.5){\line(0,1){1.2}}
\put(2.8,0.5){\line(0,1){1.2}}
\put(0.3,0.5){\line(1,1){1.1}}
\put(1.6,0.5){\line(1,1){1.1}}
\put(0.3,1.7){\line(2,-1){2.4}}
\put(0,0){$v_{82}$}
\put(1.2,0){$v_{61}$}
\put(2.4,0){$v_{52}$}
\put(0,2){$v_{20}$}
\put(1.2,2){$v_{21}$}
\put(2.4,2){$v_{22}$}
\end{picture}  
\begin{picture}(4,3)
\thicklines
\put(0.2,0.5){\line(0,1){1.2}}
\put(1.5,0.5){\line(0,1){1.2}}
\put(2.8,0.5){\line(0,1){1.2}}
\put(0.3,0.5){\line(1,1){1.1}}
\put(1.6,0.5){\line(1,1){1.1}}
\put(0.3,1.7){\line(2,-1){2.4}}
\put(0,0){$v_{60}$}
\put(1.2,0){$v_{10}$}
\put(2.4,0){$v_{50}$}
\put(0,2){$v_{30}$}
\put(1.2,2){$v_{31}$}
\put(2.4,2){$v_{32}$}
\end{picture}  
\begin{picture}(4,3)
\thicklines
\put(0.2,0.5){\line(0,1){1.2}}
\put(1.5,0.5){\line(0,1){1.2}}
\put(2.8,0.5){\line(0,1){1.2}}
\put(0.3,0.5){\line(1,1){1.1}}
\put(1.6,0.5){\line(1,1){1.1}}
\put(0.3,1.7){\line(2,-1){2.4}}
\put(0,0){$v_{12}$}
\put(1.2,0){$v_{80}$}
\put(2.4,0){$v_{51}$}
\put(0,2){$v_{40}$}
\put(1.2,2){$v_{41}$}
\put(2.4,2){$v_{42}$}
\end{picture}  
$$
$$
\begin{picture}(4,3)
\thicklines
\put(0.2,0.5){\line(0,1){1.2}}
\put(1.5,0.5){\line(0,1){1.2}}
\put(2.8,0.5){\line(0,1){1.2}}
\put(0.3,0.5){\line(1,1){1.1}}
\put(1.6,0.5){\line(1,1){1.1}}
\put(0.3,1.7){\line(2,-1){2.4}}
\put(-1.5,2){$A:$}
\put(-1.5,0){$B:$}
\put(0,0){$v_{21}$}
\put(1.2,0){$v_{31}$}
\put(2.4,0){$v_{41}$}
\put(0,2){$v_{50}$}
\put(1.2,2){$v_{51}$}
\put(2.4,2){$v_{52}$}
\end{picture}  
\begin{picture}(4,3)
\thicklines
\put(0.2,0.5){\line(0,1){1.2}}
\put(1.5,0.5){\line(0,1){1.2}}
\put(2.8,0.5){\line(0,1){1.2}}
\put(0.3,0.5){\line(1,1){1.1}}
\put(1.6,0.5){\line(1,1){1.1}}
\put(0.3,1.7){\line(2,-1){2.4}}
\put(0,0){$v_{71}$}
\put(1.2,0){$v_{32}$}
\put(2.4,0){$v_{20}$}
\put(0,2){$v_{60}$}
\put(1.2,2){$v_{61}$}
\put(2.4,2){$v_{62}$}
\end{picture}  
\begin{picture}(4,3)
\thicklines
\put(0.2,0.5){\line(0,1){1.2}}
\put(1.5,0.5){\line(0,1){1.2}}
\put(2.8,0.5){\line(0,1){1.2}}
\put(0.3,0.5){\line(1,1){1.1}}
\put(1.6,0.5){\line(1,1){1.1}}
\put(0.3,1.7){\line(2,-1){2.4}}
\put(0,0){$v_{81}$}
\put(1.2,0){$v_{11}$}
\put(2.4,0){$v_{62}$}
\put(0,2){$v_{70}$}
\put(1.2,2){$v_{71}$}
\put(2.4,2){$v_{72}$}
\end{picture}  
\begin{picture}(4,3)
\thicklines
\put(0.2,0.5){\line(0,1){1.2}}
\put(1.5,0.5){\line(0,1){1.2}}
\put(2.8,0.5){\line(0,1){1.2}}
\put(0.3,0.5){\line(1,1){1.1}}
\put(1.6,0.5){\line(1,1){1.1}}
\put(0.3,1.7){\line(2,-1){2.4}}
\put(0,0){$v_{22}$}
\put(1.2,0){$v_{40}$}
\put(2.4,0){$v_{72}$}
\put(0,2){$v_{80}$}
\put(1.2,2){$v_{81}$}
\put(2.4,2){$v_{82}$}
\end{picture}  
$$

\vspace{0,3cm}
 \underline{$B=12$}: the pair of orbits containing the vectors $(v\otimes v_{40},v\otimes v_{72})$
 \setlength{\unitlength}{0.8cm}
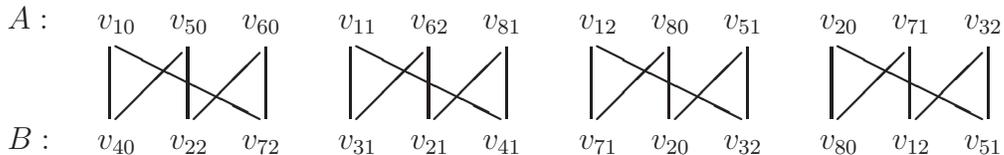
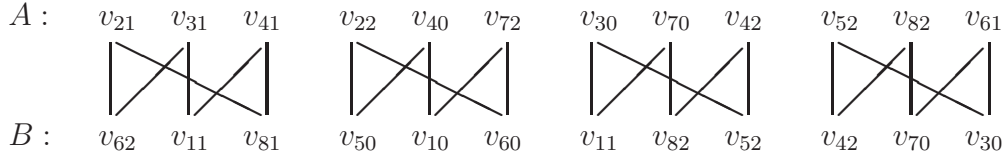
$$
\begin{picture}(4,3)
\thicklines
\put(0.2,0.5){\line(0,1){1.2}}
\put(1.5,0.5){\line(0,1){1.2}}
\put(2.8,0.5){\line(0,1){1.2}}
\put(0.3,0.5){\line(1,1){1.1}}
\put(1.6,0.5){\line(1,1){1.1}}
\put(0.3,1.7){\line(2,-1){2.4}}
\put(-1.5,2){$A:$}
\put(-1.5,0){$B:$}
\put(0,0){$v_{40}$}
\put(1.2,0){$v_{22}$}
\put(2.4,0){$v_{72}$}
\put(0,2){$v_{10}$}
\put(1.2,2){$v_{50}$}
\put(2.4,2){$v_{60}$}
\end{picture}  
\begin{picture}(4,3)
\thicklines
\put(0.2,0.5){\line(0,1){1.2}}
\put(1.5,0.5){\line(0,1){1.2}}
\put(2.8,0.5){\line(0,1){1.2}}
\put(0.3,0.5){\line(1,1){1.1}}
\put(1.6,0.5){\line(1,1){1.1}}
\put(0.3,1.7){\line(2,-1){2.4}}
\put(0,0){$v_{31}$}
\put(1.2,0){$v_{21}$}
\put(2.4,0){$v_{41}$}
\put(0,2){$v_{11}$}
\put(1.2,2){$v_{62}$}
\put(2.4,2){$v_{81}$}
\end{picture}  
\begin{picture}(4,3)
\thicklines
\put(0.2,0.5){\line(0,1){1.2}}
\put(1.5,0.5){\line(0,1){1.2}}
\put(2.8,0.5){\line(0,1){1.2}}
\put(0.3,0.5){\line(1,1){1.1}}
\put(1.6,0.5){\line(1,1){1.1}}
\put(0.3,1.7){\line(2,-1){2.4}}
\put(0,0){$v_{71}$}
\put(1.2,0){$v_{20}$}
\put(2.4,0){$v_{32}$}
\put(0,2){$v_{12}$}
\put(1.2,2){$v_{80}$}
\put(2.4,2){$v_{51}$}
\end{picture}  
\begin{picture}(4,3)
\thicklines
\put(0.2,0.5){\line(0,1){1.2}}
\put(1.5,0.5){\line(0,1){1.2}}
\put(2.8,0.5){\line(0,1){1.2}}
\put(0.3,0.5){\line(1,1){1.1}}
\put(1.6,0.5){\line(1,1){1.1}}
\put(0.3,1.7){\line(2,-1){2.4}}
\put(0,0){$v_{80}$}
\put(1.2,0){$v_{12}$}
\put(2.4,0){$v_{51}$}
\put(0,2){$v_{20}$}
\put(1.2,2){$v_{71}$}
\put(2.4,2){$v_{32}$}
\end{picture}  
$$
$$
\begin{picture}(4,3)
\thicklines
\put(0.2,0.5){\line(0,1){1.2}}
\put(1.5,0.5){\line(0,1){1.2}}
\put(2.8,0.5){\line(0,1){1.2}}
\put(0.3,0.5){\line(1,1){1.1}}
\put(1.6,0.5){\line(1,1){1.1}}
\put(0.3,1.7){\line(2,-1){2.4}}
\put(-1.5,2){$A:$}
\put(-1.5,0){$B:$}
\put(0,0){$v_{62}$}
\put(1.2,0){$v_{11}$}
\put(2.4,0){$v_{81}$}
\put(0,2){$v_{21}$}
\put(1.2,2){$v_{31}$}
\put(2.4,2){$v_{41}$}
\end{picture}  
\begin{picture}(4,3)
\thicklines
\put(0.2,0.5){\line(0,1){1.2}}
\put(1.5,0.5){\line(0,1){1.2}}
\put(2.8,0.5){\line(0,1){1.2}}
\put(0.3,0.5){\line(1,1){1.1}}
\put(1.6,0.5){\line(1,1){1.1}}
\put(0.3,1.7){\line(2,-1){2.4}}
\put(0,0){$v_{50}$}
\put(1.2,0){$v_{10}$}
\put(2.4,0){$v_{60}$}
\put(0,2){$v_{22}$}
\put(1.2,2){$v_{40}$}
\put(2.4,2){$v_{72}$}
\end{picture}  
\begin{picture}(4,3)
\thicklines
\put(0.2,0.5){\line(0,1){1.2}}
\put(1.5,0.5){\line(0,1){1.2}}
\put(2.8,0.5){\line(0,1){1.2}}
\put(0.3,0.5){\line(1,1){1.1}}
\put(1.6,0.5){\line(1,1){1.1}}
\put(0.3,1.7){\line(2,-1){2.4}}
\put(0,0){$v_{11}$}
\put(1.2,0){$v_{82}$}
\put(2.4,0){$v_{52}$}
\put(0,2){$v_{30}$}
\put(1.2,2){$v_{70}$}
\put(2.4,2){$v_{42}$}
\end{picture}  
\begin{picture}(4,3)
\thicklines
\put(0.2,0.5){\line(0,1){1.2}}
\put(1.5,0.5){\line(0,1){1.2}}
\put(2.8,0.5){\line(0,1){1.2}}
\put(0.3,0.5){\line(1,1){1.1}}
\put(1.6,0.5){\line(1,1){1.1}}
\put(0.3,1.7){\line(2,-1){2.4}}
\put(0,0){$v_{42}$}
\put(1.2,0){$v_{70}$}
\put(2.4,0){$v_{30}$}
\put(0,2){$v_{52}$}
\put(1.2,2){$v_{82}$}
\put(2.4,2){$v_{61}$}
\end{picture}  
$$

 \vspace{0,3cm}
  \underline{$B=14$}: the pair of orbits containing the vectors $(v\otimes v_{20},v\otimes v_{32})$
 \setlength{\unitlength}{0.8cm}
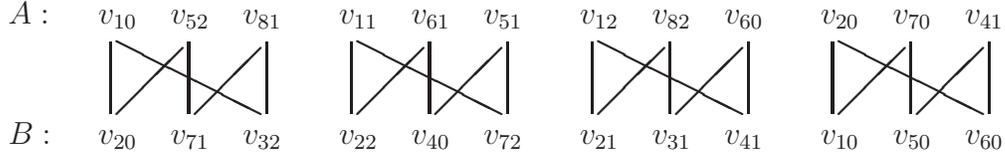
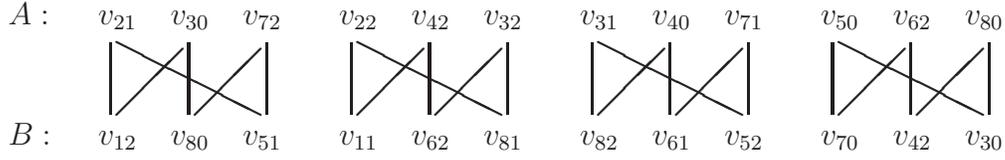
$$
\begin{picture}(4,3)
\thicklines
\put(0.2,0.5){\line(0,1){1.2}}
\put(1.5,0.5){\line(0,1){1.2}}
\put(2.8,0.5){\line(0,1){1.2}}
\put(0.3,0.5){\line(1,1){1.1}}
\put(1.6,0.5){\line(1,1){1.1}}
\put(0.3,1.7){\line(2,-1){2.4}}
\put(-1.5,2){$A:$}
\put(-1.5,0){$B:$}
\put(0,0){$v_{20}$}
\put(1.2,0){$v_{71}$}
\put(2.4,0){$v_{32}$}
\put(0,2){$v_{10}$}
\put(1.2,2){$v_{52}$}
\put(2.4,2){$v_{81}$}
\end{picture}  
\begin{picture}(4,3)
\thicklines
\put(0.2,0.5){\line(0,1){1.2}}
\put(1.5,0.5){\line(0,1){1.2}}
\put(2.8,0.5){\line(0,1){1.2}}
\put(0.3,0.5){\line(1,1){1.1}}
\put(1.6,0.5){\line(1,1){1.1}}
\put(0.3,1.7){\line(2,-1){2.4}}
\put(0,0){$v_{22}$}
\put(1.2,0){$v_{40}$}
\put(2.4,0){$v_{72}$}
\put(0,2){$v_{11}$}
\put(1.2,2){$v_{61}$}
\put(2.4,2){$v_{51}$}
\end{picture}  
\begin{picture}(4,3)
\thicklines
\put(0.2,0.5){\line(0,1){1.2}}
\put(1.5,0.5){\line(0,1){1.2}}
\put(2.8,0.5){\line(0,1){1.2}}
\put(0.3,0.5){\line(1,1){1.1}}
\put(1.6,0.5){\line(1,1){1.1}}
\put(0.3,1.7){\line(2,-1){2.4}}
\put(0,0){$v_{21}$}
\put(1.2,0){$v_{31}$}
\put(2.4,0){$v_{41}$}
\put(0,2){$v_{12}$}
\put(1.2,2){$v_{82}$}
\put(2.4,2){$v_{60}$}
\end{picture}  
\begin{picture}(4,3)
\thicklines
\put(0.2,0.5){\line(0,1){1.2}}
\put(1.5,0.5){\line(0,1){1.2}}
\put(2.8,0.5){\line(0,1){1.2}}
\put(0.3,0.5){\line(1,1){1.1}}
\put(1.6,0.5){\line(1,1){1.1}}
\put(0.3,1.7){\line(2,-1){2.4}}
\put(0,0){$v_{10}$}
\put(1.2,0){$v_{50}$}
\put(2.4,0){$v_{60}$}
\put(0,2){$v_{20}$}
\put(1.2,2){$v_{70}$}
\put(2.4,2){$v_{41}$}
\end{picture}  
$$
$$
\begin{picture}(4,3)
\thicklines
\put(0.2,0.5){\line(0,1){1.2}}
\put(1.5,0.5){\line(0,1){1.2}}
\put(2.8,0.5){\line(0,1){1.2}}
\put(0.3,0.5){\line(1,1){1.1}}
\put(1.6,0.5){\line(1,1){1.1}}
\put(0.3,1.7){\line(2,-1){2.4}}
\put(-1.5,2){$A:$}
\put(-1.5,0){$B:$}
\put(0,0){$v_{12}$}
\put(1.2,0){$v_{80}$}
\put(2.4,0){$v_{51}$}
\put(0,2){$v_{21}$}
\put(1.2,2){$v_{30}$}
\put(2.4,2){$v_{72}$}
\end{picture}  
\begin{picture}(4,3)
\thicklines
\put(0.2,0.5){\line(0,1){1.2}}
\put(1.5,0.5){\line(0,1){1.2}}
\put(2.8,0.5){\line(0,1){1.2}}
\put(0.3,0.5){\line(1,1){1.1}}
\put(1.6,0.5){\line(1,1){1.1}}
\put(0.3,1.7){\line(2,-1){2.4}}
\put(0,0){$v_{11}$}
\put(1.2,0){$v_{62}$}
\put(2.4,0){$v_{81}$}
\put(0,2){$v_{22}$}
\put(1.2,2){$v_{42}$}
\put(2.4,2){$v_{32}$}
\end{picture}  
\begin{picture}(4,3)
\thicklines
\put(0.2,0.5){\line(0,1){1.2}}
\put(1.5,0.5){\line(0,1){1.2}}
\put(2.8,0.5){\line(0,1){1.2}}
\put(0.3,0.5){\line(1,1){1.1}}
\put(1.6,0.5){\line(1,1){1.1}}
\put(0.3,1.7){\line(2,-1){2.4}}
\put(0,0){$v_{82}$}
\put(1.2,0){$v_{61}$}
\put(2.4,0){$v_{52}$}
\put(0,2){$v_{31}$}
\put(1.2,2){$v_{40}$}
\put(2.4,2){$v_{71}$}
\end{picture}  
\begin{picture}(4,3)
\thicklines
\put(0.2,0.5){\line(0,1){1.2}}
\put(1.5,0.5){\line(0,1){1.2}}
\put(2.8,0.5){\line(0,1){1.2}}
\put(0.3,0.5){\line(1,1){1.1}}
\put(1.6,0.5){\line(1,1){1.1}}
\put(0.3,1.7){\line(2,-1){2.4}}
\put(0,0){$v_{70}$}
\put(1.2,0){$v_{42}$}
\put(2.4,0){$v_{30}$}
\put(0,2){$v_{50}$}
\put(1.2,2){$v_{62}$}
\put(2.4,2){$v_{80}$}
\end{picture}  
$$

\vspace{0,3cm}
\underline{$B=16$}: the pair of orbits containing the vectors $(v\otimes v_{62},v\otimes v_{82})$
 \setlength{\unitlength}{0.8cm}
$$
\begin{picture}(3,3)
\thicklines
\put(0.2,0.5){\line(0,1){1.2}}
\put(1.5,0.5){\line(0,1){1.2}}
\put(0.3,0.5){\line(1,1){1.1}}
\put(0.3,1.7){\line(1,-1){1.1}}
\put(-1.5,2){$A:$}
\put(-1.5,0){$B:$}
\put(0,0){$v_{62}$}
\put(1.2,0){$v_{82}$}
\put(0,2){$v_{10}$}
\put(1.2,2){$v_{51}$}
\end{picture}
\begin{picture}(3,3)
\thicklines
\put(0.2,0.5){\line(0,1){1.2}}
\put(1.5,0.5){\line(0,1){1.2}}
\put(0.3,0.5){\line(1,1){1.1}}
\put(0.3,1.7){\line(1,-1){1.1}}
\put(0,0){$v_{80}$}
\put(1.2,0){$v_{52}$}
\put(0,2){$v_{11}$}
\put(1.2,2){$v_{60}$}
\end{picture}
\begin{picture}(3,2)
\thicklines
\put(0.2,0.5){\line(0,1){1.2}}
\put(1.5,0.5){\line(0,1){1.2}}
\put(0.3,0.5){\line(1,1){1.1}}
\put(0.3,1.7){\line(1,-1){1.1}}
\put(0,0){$v_{50}$}
\put(1.2,0){$v_{61}$}
\put(0,2){$v_{12}$}
\put(1.2,2){$v_{81}$}
\end{picture}
\begin{picture}(3,2)
\thicklines
\put(0.2,0.5){\line(0,1){1.2}}
\put(1.5,0.5){\line(0,1){1.2}}
\put(0.3,0.5){\line(1,1){1.1}}
\put(0.3,1.7){\line(1,-1){1.1}}
\put(0,0){$v_{31}$}
\put(1.2,0){$v_{42}$}
\put(0,2){$v_{20}$}
\put(1.2,2){$v_{72}$}
\end{picture} $$
$$ 
\begin{picture}(3,3)
\thicklines
\put(0.2,0.5){\line(0,1){1.2}}
\put(1.5,0.5){\line(0,1){1.2}}
\put(0.3,0.5){\line(1,1){1.1}}
\put(0.3,1.7){\line(1,-1){1.1}}
\put(-1.5,2){$A:$}
\put(-1.5,0){$B:$}
\put(0,0){$v_{40}$}
\put(1.2,0){$v_{70}$}
\put(0,2){$v_{21}$}
\put(1.2,2){$v_{32}$}
\end{picture}
\begin{picture}(3,3)
\thicklines
\put(0.2,0.5){\line(0,1){1.2}}
\put(1.5,0.5){\line(0,1){1.2}}
\put(0.3,0.5){\line(1,1){1.1}}
\put(0.3,1.7){\line(1,-1){1.1}}
\put(0,0){$v_{71}$}
\put(1.2,0){$v_{30}$}
\put(0,2){$v_{22}$}
\put(1.2,2){$v_{41}$}
\end{picture}
\begin{picture}(3,3)
\thicklines
\put(0.2,0.5){\line(0,1){1.2}}
\put(1.5,0.5){\line(0,1){1.2}}
\put(0.3,0.5){\line(1,1){1.1}}
\put(0.3,1.7){\line(1,-1){1.1}}
\put(0,0){$v_{41}$}
\put(1.2,0){$v_{22}$}
\put(0,2){$v_{30}$}
\put(1.2,2){$v_{71}$}
\end{picture}
\begin{picture}(3,3)
\thicklines
\put(0.2,0.5){\line(0,1){1.2}}
\put(1.5,0.5){\line(0,1){1.2}}
\put(0.3,0.5){\line(1,1){1.1}}
\put(0.3,1.7){\line(1,-1){1.1}}
\put(0,0){$v_{20}$}
\put(1.2,0){$v_{72}$}
\put(0,2){$v_{31}$}
\put(1.2,2){$v_{42}$}
\end{picture}
$$
$$ 
\begin{picture}(3,3)
\thicklines
\put(0.2,0.5){\line(0,1){1.2}}
\put(1.5,0.5){\line(0,1){1.2}}
\put(0.3,0.5){\line(1,1){1.1}}
\put(0.3,1.7){\line(1,-1){1.1}}
\put(-1.5,2){$A:$}
\put(-1.5,0){$B:$}
\put(0,0){$v_{21}$}
\put(1.2,0){$v_{32}$}
\put(0,2){$v_{40}$}
\put(1.2,2){$v_{70}$}
\end{picture}
\begin{picture}(3,3)
\thicklines
\put(0.2,0.5){\line(0,1){1.2}}
\put(1.5,0.5){\line(0,1){1.2}}
\put(0.3,0.5){\line(1,1){1.1}}
\put(0.3,1.7){\line(1,-1){1.1}}
\put(0,0){$v_{12}$}
\put(1.2,0){$v_{81}$}
\put(0,2){$v_{50}$}
\put(1.2,2){$v_{61}$}
\end{picture}
\begin{picture}(3,3)
\thicklines
\put(0.2,0.5){\line(0,1){1.2}}
\put(1.5,0.5){\line(0,1){1.2}}
\put(0.3,0.5){\line(1,1){1.1}}
\put(0.3,1.7){\line(1,-1){1.1}}
\put(0,0){$v_{60}$}
\put(1.2,0){$v_{11}$}
\put(0,2){$v_{52}$}
\put(1.2,2){$v_{80}$}
\end{picture}
\begin{picture}(3,3)
\thicklines
\put(0.2,0.5){\line(0,1){1.2}}
\put(1.5,0.5){\line(0,1){1.2}}
\put(0.3,0.5){\line(1,1){1.1}}
\put(0.3,1.7){\line(1,-1){1.1}}
\put(0,0){$v_{10}$}
\put(1.2,0){$v_{51}$}
\put(0,2){$v_{62}$}
\put(1.2,2){$v_{82}$}
\end{picture}
$$

 \begin{table}[ht!]
 \centering
 \caption{The eigenvalues of $X(\tilde{g},v)$ for all elements $\tilde{g}\in S_4$}
 \begin{tabular}{|c|c|c|c|c|c|c|c|}
 \hline
 Equivalence & $\tilde{g}\equiv g_\alpha g^l$ & $(\alpha,l)$ & \multicolumn{5}{|c|}{Eigenvalues of $X(\tilde{g},v)$}\\
 \cline{4-8}
 class & & & $D$ & $\tilde{D}$ & $D_2$ & $D_0$ & $x_{max}$\\
 \hline
  & $(1342)$ & (5,0) & 4.63 & 1.35 & 0.38 & 5.30 & 5.30\\
  \cline{2-8}
  & (1423) & (6,0) & 4.63 & 1.35 & 0.38 & 5.30 & 5.30\\
  \cline{2-8}
  & (2314) & (1,1) & 1.98 & 4.00 & 3.03 & 0.00 & 4.00\\
  \cline{2-8}
I class  & (2431) & (6,1) & 2.37 & 3.60 & 2.64 & 0.79 & 3.60\\
  \cline{2-8}
(8 elements)  & (3124) & (1,2) & 1.98 & 4.00 & 3.03 & 0.00 & 4.00\\
  \cline{2-8}
   & (3241) & (5,2) & 2.98 & 3.00 & 2.04 & 1.99 & 3.00\\
  \cline{2-8}
  & (4132) & (8,0) & 2.37 & 3.60 & 2.64 & 0.79 & 3.60\\
  \cline{2-8}
  & (4213) & (8,1) & 2.98 & 3.00 & 2.04 & 1.99 & 3.00\\
  \hline
  & (1243) & (7,2) & 3.95 & 0.30 & 1.92 & 7.40 & 7.40\\
  \cline{2-8}
  & (1324) & (2,2) & 4.60 & 0.68 & 0.77 & 6.63 & 6.63\\
  \cline{2-8}
II class  & (1432) & (4,0) & 4.76 & 1.71 & 0.00 & 4.57 & 4.76\\
  \cline{2-8}
  (6 elements)& (2134) & (2,0) & 0.69 & 3.56 & 5.18 & 0.89 & 5.18\\
  \cline{2-8}
  & (3214) & (2,1) & 2.71 & 3.76 & 2.05 & 0.48 & 3.76\\
  \cline{2-8}
  & (4231) & (3,0) & 3.33 & 1.94 & 2.03 & 4.12 & 4.12 \\
  \hline
  & (2341) & (7,0) & 2.74 & 3.74 & 2.02 & 0.52 & 3.74\\
  \cline{2-8}
  & (2413) & (4,1) & 1.31 & 3.96 & 4.05 & 0.07 & 4.05\\
  \cline{2-8}
III class  & (3421) & (4,2) & 1.93 & 2.32 & 3.95 & 3.30 & 3.95\\
  \cline{2-8}
 (6 elements) & (3142) & (7,1) & 1.31 & 3.96 & 4.05 & 0.07 & 4.05\\
  \cline{2-8}
  & (4312) & (3,1) & 1.92 & 2.32 & 3.95 & 3.36 & 3.95\\
  \cline{2-8}
  & (4123) & (3,2) & 2.74 & 3.74 & 2.02 & 0.52 & 3.74\\
  \hline
IV class  & (2143) & (5,1) & 0.40 & 3.65 & 5.58 & 0.70 & 5.58\\
  \cline{2-8}
  (3 elements) & (3412) & (6,2) & 0.99 & 3.05 & 4.98 & 1.90 & 4.98\\
  \cline{2-8}
  & (4321) & (8,2) & 2.65 & 1.39 & 3.33 & 5.21 & 5.21\\
  \hline
  V class & (1234) & (1,0) & 4.05 & 0.00 & 1.93 & 8.00 & 8.00\\
  (1 element) & & &&&&&\\
  \hline
 \end{tabular}\label{t2}
 \end{table}

\end{document}